\documentclass[showpacs,amsmath,amssymb,prd]{revtex4}
\usepackage{epsfig}
\usepackage{graphicx}% Include figure files
\usepackage{dcolumn}% Align table columns on decimal point
\usepackage{bm}% bold math
\begin{document}
%title
\title{\boldmath First observations of $\psi(2S)$ and $\chi_{cJ}(1P)$ decays
to four-body final states $h^+h^- K^0_S K^0_S$
\footnote[4]{The $h^\pm$ denote charged pions or kaons.}
}
\author{
M.~Ablikim$^{1}$, J.~Z.~Bai$^{1}$, Y.~Ban$^{10}$,
J.~G.~Bian$^{1}$, X.~Cai$^{1}$, J.~F.~Chang$^{1}$,
H.~F.~Chen$^{16}$, H.~S.~Chen$^{1}$, H.~X.~Chen$^{1}$,
J.~C.~Chen$^{1}$
% ...
\\(BES Collaboration)\\
}
\affiliation{
$^1$ Institute of High Energy Physics, Beijing 100039, People's Republic of
China\\
% ...
$^{10}$ Peking University, Beijing 100871, People's Republic of China\\
% ...
$^{16}$ University of Science and Technology of China, Hefei 230026, People's
Republic of China\\
% ...
}
\begin{abstract}
First observations of
$\chi_{c0}$, $\chi_{c1}$, and $\chi_{c2}$ decays
to $\pi^+\pi^-K^0_SK^0_S$ and $K^+K^-K^0_SK^0_S$, as well as
$\psi(2S)$ decay to $\pi^+\pi^-K^0_SK^0_S$
are presented. The branching fractions of these decay channels are
determined using $14\times 10^6$ $\psi(2S)$ events collected at BESII/BEPC.
The branching fractions of $\chi_{c0},\chi_{c2}\rightarrow K^0_SK^0_S$ are
measured with improved statistical precision.
\end{abstract}
\maketitle

\renewcommand{\textfloatsep}{0.3cm}
\renewcommand{\dbltextfloatsep}{0.3cm}

\section{Introduction}
Experimental data on charmonia and their decay properties are
essential input to test QCD models and QCD based calculations.
The importance of the Color Octet Mechanism(COM)~\cite{BBL}
in radiative decays of the $\Upsilon$~\cite{Maltoni},
$J/\psi$ production in inclusive B decays~\cite{Beneke}, as well
as inclusive decays of P-wave charmonia~\cite{KT Chao} has been
emphasized for many years.
Recently, QCD predictions of two-body exclusive decays of P-wave
charmonium with the inclusion of the COM have been 
made~\cite{JBolz,SMH Wong}
and compared to previous measurements~\cite{BES chic2pp lambda,PDG}.
More experimental data of two- and four-body exclusive decays of P-wave
charmonia with improved precision are important
for further testing this new QCD approach including the effect of the COM.

In this paper, results on $\psi(2S)$ and $\chi_{cJ}$~($J=0,1,2$)
two- and four-body hadronic decays with inclusion of a pair of 
$K^0_S$ mesons are presented. This analysis is based on $14\times10^6$
$\psi(2S)$ decays collected with BESII at the BEPC $e^+e^-$ Collider.
A sample of 6.42 pb$^{-1}$ data taken at 3.65 GeV is used for continuum 
background studies.

\section{BES Detector}
The BESII detector is described elsewhere~\cite{besii_detector}.
Charged particle momenta are determined with a resolution of $\sigma
_p/p = 1.78\% \sqrt{1+p^2}$ ($p$ in $\hbox{\rm GeV}/c$) in a 40-layer
main drift chamber (MDC).  Particle identification is
accomplished using specific ionization ($dE/dx$) information in the
drift chamber and time-of-flight (TOF) information in a barrel-like
array of 48 scintillation counters. The $dE/dx$ resolution is
$\sigma_{dE/dx}= 8\%$; the TOF resolution is $\sigma_{TOF}$=200 ps for
hadrons.  A 12-radiation-length barrel shower counter (BSC) measures
energies of photons with a resolution of $\sigma_E/E=21\%/\sqrt{E}$
($E$ in GeV).

\section{Monte Carlo Simulation}
A Geant3 based Monte Carlo, SIMBES~\cite{simbes}, which simulates the
detector response, including interactions of secondary particles in
the detector material, is used to determine detection efficiencies
and mass resolutions, as well as to optimize selection criteria and 
estimate backgrounds. Under the assumption of a pure E1 transition,
the distribution of polar angle $\theta$ of the photon in
$\psi(2S)\rightarrow \gamma\chi_{cJ}$ decays is given by
$1+k\cos^2\theta$~\cite{karl} with $k = 1,-\frac{1}{3}$, and 
$\frac{1}{13}$ for $J=0,1$, and $2$, respectively.
The angular distributions for $K^0_S$ mesons from 
$\chi_{c0,2} \rightarrow K^0_SK^0_S$ decays
are produced according to the model of $\chi_{cJ} \rightarrow
M\bar{M}$~\cite{kabir}, where $M$ stands for a $0^-$ meson.
Angular distributions for daughters from other decays
are generated isotropically in the center-of-mass system of
the $\psi(2S)$ or $\chi_{cJ}$.

\section{Data Analysis}
To be regarded as a good photon, a shower cluster in the BSC must have
an energy deposit of more than 50 MeV and at least one hit in the
first six layers of the BSC. To remove soft photons emitted by charged
particles, the differences of azimuthal angles, $d\phi$, and $z$
coordinates at the first layer of the BSC, $dz$,  between good photons and
each
charged track must satisfy either a loose requirement (selection-A:
$d\phi>10^{\circ}$ or $dz>0.3$ m)  or a tight requirement (selection-B:
$d\phi>20^{\circ}$ or $dz>1.0$ m).  Here the $z$ coordinate is defined
to point in the positron direction.

Each charged track is required to have a good helix fit. For final
states containing charged kaons, particle identification is required;
usable particle identification information in one or both of the MDC
($dE/dx$) and TOF subsystems is necessary.  A particle identification
$\chi^2$ is calculated for each track for the pion, kaon or proton
hypotheses using this information, and the associated probability $prob$
is determined. A track is identified as a kaon, if
the probability of the track being a kaon $prob(K) >0.01$; otherwise
it is regarded as a pion.  For final states containing only pions, no
particle identification is done and all tracks are assumed to be
pions.

Each event is required to contain two $K^0_S$
mesons. The reconstruction of the decay $K^0_S\rightarrow\pi^+\pi^-$
and related checks are described in detail elsewhere~\cite{K0s rec}.
A $K^0_S$ candidate must satisfy $|M_{\pi^+\pi^-}-M_{K^0_S}|
< 20$ MeV and have a decay length transverse to the beam axis $R_{xy}
> 0.3$ cm.
The $K^0_S$ sideband sample, used for background estimation,
is selected with one $\pi^+\pi^-$ pair
within the $K^0_S$ mass window and the other pair
in the $K^0_S$ mass sideband region defined by
40 MeV $<|M_{\pi^+\pi^-}-M_{K^0_S}|<$ 60 MeV.

Four constraint (4C) kinematic fits are
performed on the selected events for the following decay modes :
(1) $\psi(2S)\rightarrow \gamma K^0_SK^0_S$,
(2) $\psi(2S)\rightarrow \gamma\pi^+\pi^-K^0_SK^0_S$, and
(3) $\psi(2S)\rightarrow \gamma K^+K^-K^0_SK^0_S$.
The fits are made to each combination of a good photon and two
$K^0_S$ candidates in an event,
the combination with the minimum $\chi^2_{4C}$ is selected, and the
$\chi^2_{4C}$ is required to be less than 35. The associated probability
$prob_{4C}$ is calculated.

Background from $\psi(2S)\rightarrow \pi^+\pi^-J/\psi$ decay is
removed by calculating the mass recoiling, $M_{recoil}$, against all
pairs of oppositely charged tracks, assuming them to be pions, and
requiring $|M_{recoil}-M_{J/\psi}| > 25$ MeV.
Background contamination from
continuum production is found to be negligible for all decay
channels.

An unbinned maximum likelihood method is used in fitting the signal
for all decay channels except $\psi(2S)\rightarrow h^+h^-K^0_SK^0_S$.
The branching fractions of
$\psi(2S)\rightarrow \gamma\chi_{cJ}$~($J=0,1,2$) needed in the measurement
are taken from Particle Data Group (PDG) tables~\cite{PDG}.

\subsection{$\psi(2S)\rightarrow \gamma K^0_SK^0_S$}
\begin{figure}[h]
\psfig{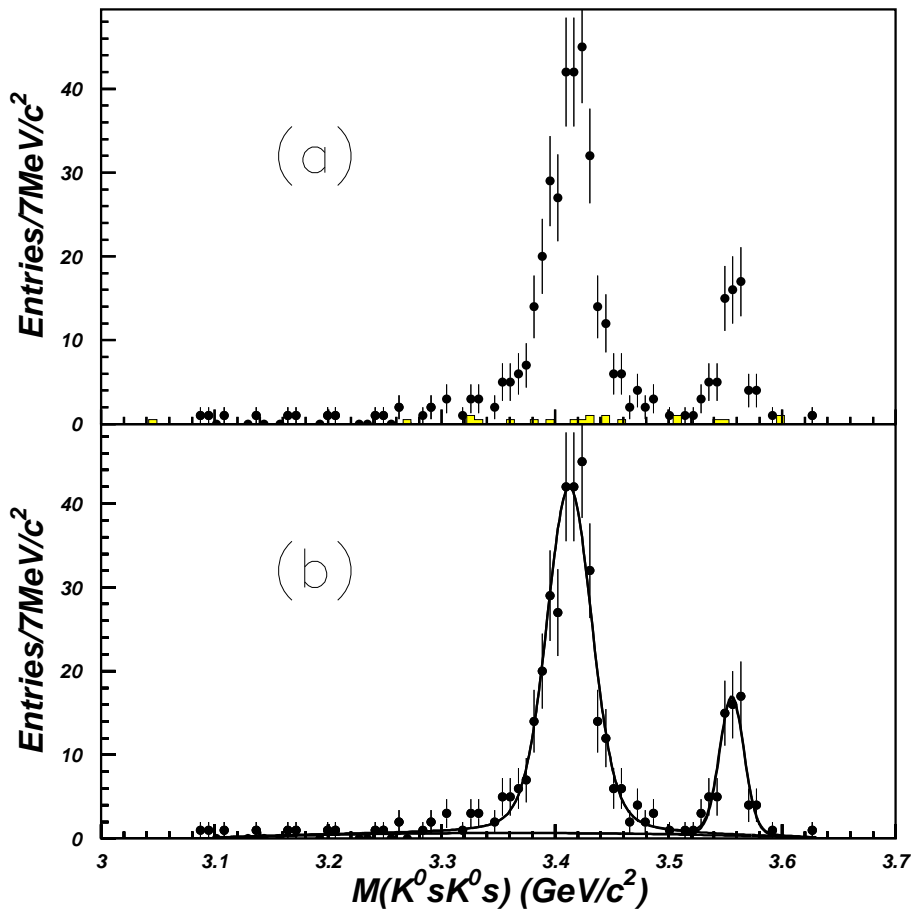}
\caption{Distribution of $K^0_SK^0_S$ invariant mass of
$\psi(2S)\rightarrow \gamma K^0_SK^0_S$ candidates. 
(a) Points with error bars are data, and the histogram is sideband background.
(b) Points with error bars are data, and the solid line is the fit described in the text.}
\label{fig:mksks}
\end{figure}

The decay $\psi(2S)\rightarrow \gamma K^0_SK^0_S$ has one photon plus
a pair of $K^0_S$ candidates. The event should have four charged
tracks with total charge zero. The loose photon selection,
selection-A, is applied because of the low background in the
channel.  The $K^0_SK^0_S$ invariant mass distribution of the selected
events is shown in Fig.~\ref{fig:mksks}.
A few $K^0_S$ sideband events survive
the selection, which is consistent with the low background
observed in Fig.~\ref{fig:mksks} (a).
No background is expected from $\psi(2S)\rightarrow \gamma \chi_{cJ}$ with
$\chi_{cJ}\rightarrow 2(\pi^+\pi^-)$ for $J=0,1,2$ and
$\psi(2S)\rightarrow \gamma \chi_{c1}$ with $\chi_{c1}\rightarrow
K^0_SK^{\pm}\pi^{\mp}$ according to the analysis of simulated MC events.

The $K^0_SK^0_S$ invariant mass distribution is fitted with
two Breit-Wigner resonances for $\chi_{c0}$ and $\chi_{c2}$, each convoluted
with Gaussian resolution functions, plus a second order polynomial background.
The $\chi_{c0,2}$ widths in the fitting are fixed to their PDG values~\cite{PDG}.
The resulting fit is shown in Fig.~\ref{fig:mksks} (b).
Including the $\chi_{c1}$ resonance in the fit
yields zero events for the CP violating decay
$\chi_{c1}\rightarrow K^0_SK^0_S$.

\subsection{$\psi(2S)\rightarrow \gamma\pi^+\pi^-K^0_SK^0_S$}
\begin{figure}[h]
\psfig{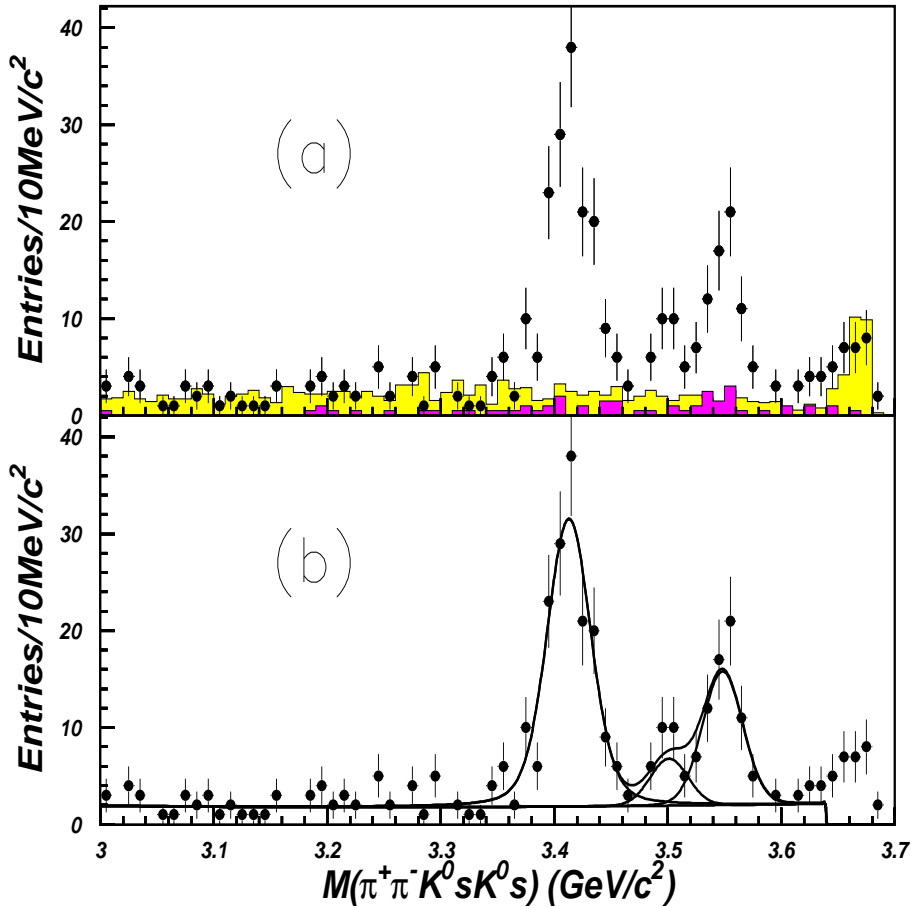}
\caption{Distribution of $\pi^+\pi^-K^0_SK^0_S$ invariant mass for
$\psi(2S)\rightarrow \gamma\pi^+\pi^-K^0_SK^0_S$ candidates.
Points with error bars are data.
The light shaded area in (a) is background simulation, where 
some unknown branching ratios are 
normalized to agree with the overall $\chi_{cJ}$ background level,
and the dark shaded area is $K^0_S$ sideband.
The solid line in (b) is the fit.}
\label{fig:mpipiksks}
\end{figure}

The $\psi(2S)\rightarrow \gamma\pi^+\pi^-K^0_SK^0_S$
decay channel contains one photon and six charged tracks with total
charge zero.  The requirements here are similar to the previous case,
but there are two additional pions.
Background from $\pi/K$ misidentification is suppressed by the
requirement $prob_{4C}(\gamma\pi^+\pi^-K^0_SK^0_S)
>prob_{4C}(\gamma K^+K^-K^0_SK^0_S)$.
The $\pi^+\pi^-K^0_SK^0_S$ invariant mass distribution for selected events
is shown in Fig. \ref{fig:mpipiksks}.

In Fig. \ref{fig:mpipiksks} there are two kinds of background in the mass region between 3.0 and
3.64 GeV/c$^2$: (1) background corresponding to $K_S^0$ sidebands, and
(2) $\psi(2S)$ decays and $\chi_{cJ}$ decays different from the
signal channel, where the decays also include a pair of $K^0_S$ mesons.
Studies with $K^0_S$ sideband events for both data and MC
show that $K^0_S$ sideband background from wrong
combinations of $\pi^+\pi^-$ is slightly enhanced in the 
$\chi_{cJ}$ signal region.  MC studies show that the smooth 
background spread over the whole mass region from (2)
results mainly from the following decay channels: 
(a)~$\psi(2S)\rightarrow
\gamma \chi_{cJ}$ with $\chi_{cJ}\rightarrow 3(\pi^+\pi^-)$ and
$\chi_{cJ}\rightarrow K^+K^-K^0_SK^0_S$, (b)~$\psi(2S)\rightarrow
\pi^0\pi^+\pi^-K^0_SK^0_S$, and
(c)~$\psi(2S)\rightarrow \omega K^0_SK^0_S$ with $\omega\rightarrow
\pi^+\pi^-\pi^0$.  Background events in the high mass region above
$3.64$ GeV/c$^2$ in Fig. \ref{fig:mpipiksks} are from
$\psi(2S)\rightarrow \pi^+\pi^-K^0_SK^0_S$ decays combined with an
unassociated low energy photon.

%MC simulation is used to figure out the situation of background
%and they are also plotted in Fig.~\ref{fig:mpipiksks} (a) (light shaded area).
%They are mainly from:
%(a)~$\psi(2S)\rightarrow \pi^+\pi^-K^0_SK^0_S$ which contributes to
%the high end of it (above 3.64 $GeV/c^2$).
%(b)~$\psi(2S)\rightarrow \omega K^0_SK^0_S$ with $\omega\rightarrow
%\pi^+\pi^-\pi^0$ estimated from $\psi(2S)\rightarrow \omega K^+K^-$\cite{PDG}.
%(c)~$\psi(2S)\rightarrow \pi^0\pi^+\pi^-K^0_SK^0_S$ that possibly comes
%from $K^*$ gerenation in $\psi(2S)$ decay. 
%In Fig.~\ref{fig:mpipiksks} their unknown branching ratio is normalized to
%background level of data.
%Obviously they all can't form a strong enhancement in the 
%signal region around 3.5 $GeV/c^2$.

%Background from $K^0_S$ sideband event is also presented 
%in Fig.~\ref{fig:mpipiksks} (a) (dark shaded area).
%It is slight and partly from the wrong combination of $\pi^+\pi^-$ 
%which is about 2\% known from signal's simulation.
%This check tells that background events with only one $K^0_S$
%can't contaminate the signal.

The $\pi^+\pi^-K^0_SK^0_S$ invariant mass distribution between
3.0 to 3.64 GeV/c$^2$ is fitted with
three Breit-Wigner resonances $\chi_{cJ}$~($J=0,1,2$),
convoluted with Gaussian resolution functions, plus a second order
polynomial background. The widths of the $\chi_{c0,1,2}$ resonances in 
the fit are fixed to their PDG values. The fit is shown in 
Fig.~\ref{fig:mpipiksks}.
The numbers of events in the three peaks 
determined from the fit include signal and 
$K_S^0$ sideband background, which is somewhat enhanced in the regions of the
peaks. The $K_S^0$ sideband sample for data is fitted with a fake
signal shape, found by fitting the MC $K_S^0$ sideband sample,
plus a second order polynomial background.
The numbers of sideband
background events,
5.3, 0.6 and 5.5 for $\chi_{c0}$, $\chi_{c1}$ and $\chi_{c2}$
respectively, are then subtracted from the total numbers of events
in three peaks.

%The $\pi^+\pi^-K^0_SK^0_S$ invariant mass distribution from
%sideband events are also fitted. The shape parameters of $\chi_{cJ}$ 
%are got from $K^0_S$ signal events fitting and are fixed. And their
%contributions for each $\chi_{cJ}$ resonance 
%are substracted from signal calculation.

\subsection{$\psi(2S)\rightarrow \gamma K^+K^-K^0_SK^0_S$}
\begin{figure}[h]
\psfig{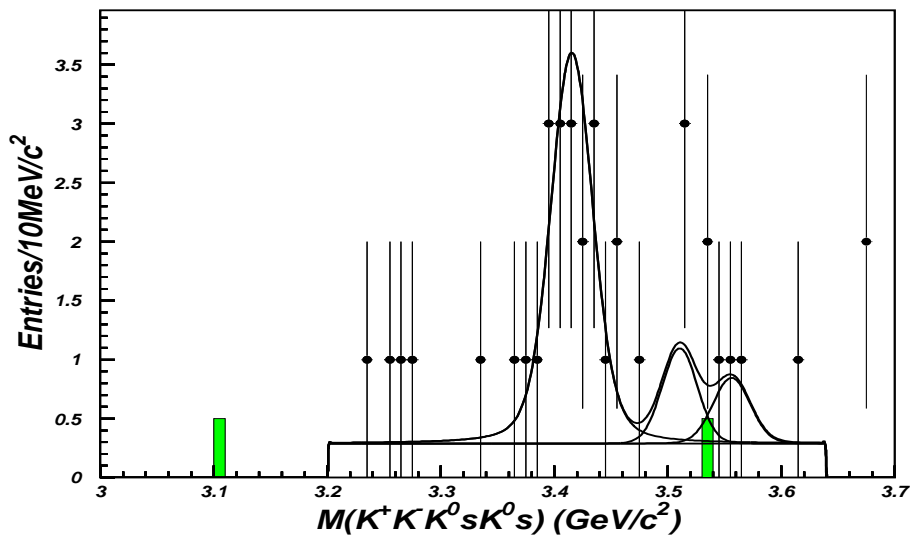}
\caption{Distribution of $K^+K^-K^0_SK^0_S$ invariant mass of
$\psi(2S)\rightarrow \gamma K^+K^-K^0_SK^0_S$ candidates. Points with
error bars are data, and the histogram is sideband background.
The solid line is the fit.}
\label{fig:mkkksks}
\end{figure}

The $\psi(2S)\rightarrow \gamma K^+K^-K^0_SK^0_S$ decay has the same
topology as $\psi(2S)\rightarrow \gamma \pi^+\pi^-K^0_SK^0_S$, and
thus it is subject to
similar event selection criteria except for the kaon
identification requirement for two of the charged tracks.  First, the
$K^0_SK^0_S$ pair is searched for under the assumption that all charged
tracks are pions. Kaon identification is only done for the two charged
tracks remaining after reconstruction of the $K^0_SK^0_S$
pair. We also require $prob_{4C}(\gamma
K^+K^-K^0_SK^0_S)>prob_{4C}(\gamma\pi^+\pi^-K^0_SK^0_S)$ for the 4C
kinematic fit probabilities to suppress contamination from
$\psi(2S)\rightarrow \gamma\pi^+\pi^-K^0_SK^0_S$ decays. The
$K^+K^-K^0_SK^0_S$ invariant mass distribution for selected events is 
shown in Fig.~\ref{fig:mkkksks}.

As seen from Fig.~\ref{fig:mkkksks} 
only one event survives from the $K^0_S$ sideband sample for data.  MC
events for the following possible  background channels are generated:
(1)~$\psi(2S)\rightarrow \gamma \chi_{cJ}$ with $\chi_{cJ}\rightarrow
3(\pi^+\pi^-)$ and $\pi^+\pi^-K^0_SK^0_S$, (2)~$\psi(2S)\rightarrow
\pi^+\pi^-K^0_SK^0_S$, and (3)~$\psi(2S)\rightarrow \omega K^0_SK^0_S$
with $\omega\rightarrow \pi^+\pi^-\pi^0$. However, no event from these
background channels %is survived.  
survives the selection criteria.
Another study with a large sample of
simulated $\psi(2S)\rightarrow anything$~\cite{chenjc} shows
that negligible background comes from decays of $\psi(2S)\rightarrow
\phi K^{*0}K^0  \rightarrow\pi^0 K^+K^- K^0_SK^0_S$.

The $K^+K^-K^0_SK^0_S$ invariant mass distribution is fitted with
three Breit-Wigner resonances, $\chi_{cJ}$ ($J=0,1,2$), convoluted
with Gaussian resolution functions, plus a flat background.
Because of low statistics in the signal region, not only the widths
and mass resolutions for the $\chi_{cJ}$ ($J=0,1,2$),
but also the masses of the $\chi_{c1}$ and $\chi_{c2}$ in the fitting
are fixed to their PDG values.
The fitting results are shown in the Fig. \ref{fig:mkkksks}.

%Section No. 2.4.4
\subsection{$\psi(2S)\rightarrow h^+h^- K^0_SK^0_S$}
\begin{figure}[h]
\psfig{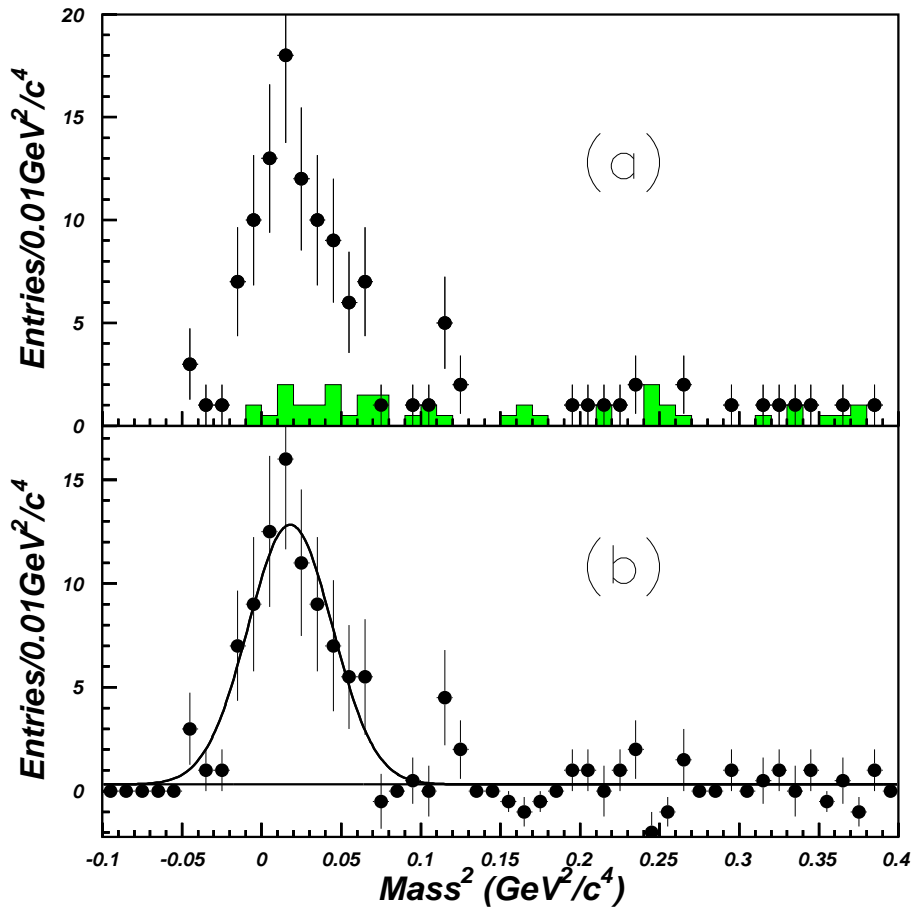}
\caption{Distribution of invariant mass squared of the two
remaining charged particles
after $K^0_SK^0_S$ selection for $\psi(2S)\rightarrow
h^+h^-K^0_SK^0_S$.  (a) Points with error bars are data.  The
histogram is the $K^0_S$ sideband background.  (b) Points with
error bars are the data with the $K^0_S$ sideband background
subtracted. The solid line is the fit.}
\label{fig:mhsquared}
\end{figure}

The selection of $\psi(2S)\rightarrow h^+h^- K^0_SK^0_S$ decays
requires six charged tracks with total charge zero and no good photon
in the event, as defined above. Good photons are rejected
with the tight selection,
selection-B, in order to gain higher detection efficiency for signal
events.  The $K^0_S$ reconstruction uses all combinations of
oppositely charged tracks assuming all tracks are pions.  To further
suppress background of $\psi(2S)$ radiative decays, a requirement on the
missing momentum of six charged tracks is employed: $P_{miss} < 80$
MeV.  The two charged tracks $h^+$ and $h^-$
recoiling against the $K^0_S$ pair are assumed to have the same mass $m$.
Using energy-momentum
conservation, the mass squared $m^2$ is calculated from
\begin{equation}
m^2=\frac{E^4+(P^2_{h^+}-P^2_{h^-})^2-2E^2(P^2_{h^+}+P^2_{h^-})}
{4E^2}
\label{for:eq}
\end{equation}
where $E=M_{\psi(2S)}-E_{K^0_SK^0_S}$, and $P_{h^\pm}$ is the momentum
of $h^+$ or $h^-$. The distribution of $m^2$ for selected events
is shown in Fig.~\ref{fig:mhsquared}.  The peak at low mass is
consistent with $\pi^+ \pi^-$; there is no evidence for $K^+ K^-$.

Two events from the continuum data sample survive the above selection
and their effect will be included in the systematic error.  No
background is found in MC studies of the following decay channels :
(1) $\psi(2S)\rightarrow \gamma \chi_{cJ}$ with $\chi_{cJ}\rightarrow
3(\pi^+\pi^-)$, $\pi^+\pi^-K^0_SK^0_S$, and $K^+K^-K^0_SK^0_S$ and (2)
$\psi(2S)\rightarrow \omega K^0_SK^0_S$ with $\omega\rightarrow
\pi^+\pi^-\pi^0$.  Background estimated using the $K^0_S$ sideband
data is subtracted from the observed number of signal events.
A MC study shows that the shape of
the charged pion signal in the $m^2$ spectrum is well described by a
Gaussian function, and its mean and resolution are consistent with data.
The spectrum is fitted with a Gaussian signal function
and a flat background using a  binned maximum likelihood fit where the
resolution is fixed to the MC determined value. The
fitting result is shown in the Fig.~\ref{fig:mhsquared}.

\subsection{Systematic Errors}
Systematic errors for the efficiency are caused by differences between
data and MC simulation.  Our studies have determined these errors to be
$2\%$ per track for the tracking efficiency, $2\%$ for photon
identification, $5\%$ for the 4C kinematic fit, and $2.1\%$ for the
$K^0_S$ reconstruction efficiency.  A correction factor due to the
overestimate of the $K^0_S$ reconstruction efficiency of the MC relative
to data is determined to be $95.8\%$.
The change of fitting range and background shape function 
contributes a difference of final results less than $3\%$.
Other systematic errors arise from the uncertainties in the total
number of $\psi(2S)$ events, $(14.00\pm0.56)\times 10^6$~\cite{moxh},
and in the branching fractions for $K^0_S\rightarrow \pi^+\pi^-$ and
$\psi(2S)\rightarrow \gamma \chi_{cJ}$ (J=0,1,2).
In $\psi(2S)\rightarrow \pi^+\pi^-K^0_SK^0_S$ decay,
with two events found in continuum data,
an additional error of $7.7\%$ is added.

\subsection{Result and Discussion}
Possible resonance structures have been
searched for the $\chi_{c0} \rightarrow \pi^+\pi^-K^0_SK^0_S$
final state which is the channel with the highest number of observed events.
Some excess for inclusive decays of $K^{*}(892)^+ \rightarrow K^0_S \pi^+$,
$f_0 (1710) \rightarrow K^0_S K^0_S$, $\rho(770)\rightarrow \pi^+\pi^-$
and $f_0 (980) \rightarrow \pi^+\pi^-$ can be seen from the
selected events.
%Insufficiecte statistics and complicate structures in these decay
%modes make us hard to reconstruct clean signal for
%two-body decay with intermediate resonances.
Insufficient statistics and complicated structures in these
decay modes make it difficult to identify clear signals for two-body
decays with intermediate resonances.
Efficiencies for final states with resonances, such as
$K^{*}(892)^+ K^{*}(892)^-$,
$K_0^{*}(1430)^+ K_0^{*}(1430)^-$,
$K_0^{*}(1430)^+ K_2^{*}(1430)^-$,
$f_0(1370) f_0(1710)$,
$f_0(980) f_0(980)$,
$f_0(980) f_0(2200)$ and $K_1(1270)^0K^0$~\cite{guo}
are studied using phase space MC events.
The averaged difference in efficiency between final states
with and without
intermediate resonance is estimated to be $7.7\%$, which is
regarded as systematic error in the measurements of the branching
fractions for the four-body final states.
The results of four-body
final states $h^+h^-K^0_S K^0_S$ in our mesurements include
those of both non-resonance and intermediate resonance.

Final results of signal yield and branching fractions for
the $\chi_{cJ}(1P)$ and
$\psi(2S)$ two- and four-body hadronic decays involving $K^0_S$ pair
production are summarized in Table~\ref{tab:fitting}. The masses
of the $\chi_{cJ}$~(J=0,1,2) extracted from the fits are also listed.
The 90\% confidence level (CL) upper limits on the branching fractions in the table are 
obtained using the Feldman-Cousins method~\cite{up}.
The branching fractions of $\chi_{cJ}$~(J=0,1,2) decays to
$\pi^+\pi^-K^0_SK^0_S$ and $K^+K^-K^0_SK^0_S$, as well $\psi(2S)$
decay to $\pi^+\pi^-K^0_SK^0_S$ are observed for the first time. The
branching fractions of $\chi_{c0}$ and $\chi_{c2}$ decays to
$K^0_SK^0_S$ are measured with improved precision.

%table for this section
\begin{table*}
\begin{center}
\caption{Summary of the fitting results. Errors for the signal yield
$n_{s}$, background $n_b$, mass $M$, and mass squared $m^2$ are
statistical. The detection efficiency $\epsilon$ and resolution $\sigma$ 
for each decay channel from MC are shown.}
\label{tab:fitting}
\begin{tabular}{|c|ccc|cc|cc|c|}\hline
{Channel} & {$n_{s}$} & {$n_b$} & {$M_{\chi_{cJ}}$} & {$\epsilon$} & {$\sigma$}\\
& & & {(MeV/c$^2$)} & {(\%)} & {(MeV/c$^2$)}\\\hline

{$\chi_{c0}\rightarrow K^0_SK^0_S$}&{$322\pm20$}&{}
&{3413.1$\pm$1.2}&{$7.96$}&{13.3}\\
{$\chi_{c1}\rightarrow K^0_SK^0_S$}&{0}&{$6.4\pm2.6$}
&{fixed}         &{$8.50$}&{12.8}\\
{$\chi_{c2}\rightarrow K^0_SK^0_S$}&{$65.1\pm8.7$}&{}
&{3555.7$\pm$1.8}&{$8.48$}&{11.8}\\\hline

{$\chi_{c0}\rightarrow \pi^+\pi^-K^0_SK^0_S$}&{$152\pm14$}&{}
&{$3412.9\pm2.0$}&{$2.03$}&{16.8}\\
{$\chi_{c1}\rightarrow \pi^+\pi^-K^0_SK^0_S$}&{$19.8\pm7.7$}&{}
&{$3501.1\pm6.2$}&{$2.20$}&{16.4}\\
{$\chi_{c2}\rightarrow \pi^+\pi^-K^0_SK^0_S$}&{$57\pm11$}&{}
&{$3548.2\pm3.1$}&{$2.04$}&{17.2}\\\hline

{$\chi_{c0}\rightarrow K^+K^-K^0_SK^0_S$}&{$16.8\pm4.8$}&{}
&{$3415.4\pm6.1$}&{$0.91$}&{16.1}\\
{$\chi_{c1}\rightarrow K^+K^-K^0_SK^0_S$}&{$3.2\pm2.4$}&{$1.8\pm0.8$}
&{fixed}         &{$1.12$}&{15.3}\\
{$\chi_{c2}\rightarrow K^+K^-K^0_SK^0_S$}&{$2.3\pm2.2$}&{$1.8\pm0.8$}
&{fixed}&{$1.05$}&{15.9}\\\hline\hline

{Channel}&{$n_{s}$} & {$n_b$}
&{$m^2$($10^{-3}$)}&{$\epsilon$}&{$\sigma$($10^{-3}$)}\\
& & &{(GeV$^2$/c$^4)$}&{(\%)}&{(GeV$^2$/c$^4)$}\\\hline
{$\psi(2S)\rightarrow \pi^+\pi^-K^0_SK^0_S$}&{$83.2\pm9.4$} & {}
&{$18.0\pm3.1$}  &{$2.82$}&{26.5}\\\hline
\end{tabular}
\end{center}
\end{table*}

%table of branching ratios
\begin{table*}
\begin{center}
\caption{The branching fractions from this measurement, as well as
previous results, are listed.
The first and second errors for the branching
fractions $BR$ are statistical and systematic, respectively.}  
\label{tab:branching ratio}
\begin{tabular}{|c|cc|c|}\hline
{Channel} & {$BR(\psi(2S)\rightarrow \gamma\chi_c)\times BR(\chi_c\rightarrow X)$}
&{$BR(\chi_c\rightarrow X)$}&{$BR_{PDG}(\chi_c\rightarrow X)$\cite{PDG}}\\
{}&{($10^{-5}$)}&{($10^{-4}$)}&{($10^{-4}$)}\\\hline

{$\chi_{c0}\rightarrow K^0_SK^0_S$}
&{$30.2\pm1.9\pm3.3$}&{$35.1\pm2.2\pm4.7$}&{$21\pm6$}\\
{$\chi_{c1}\rightarrow K^0_SK^0_S$}&{$<0.6$ (CL=90\%)}&{$<0.8$ (CL=90\%)}&{-}\\
{$\chi_{c2}\rightarrow K^0_SK^0_S$}
&{$5.72\pm0.76\pm0.63$}&{$8.9\pm1.2\pm1.3$}&{$7.2\pm2.7$}\\\hline

{$\chi_{c0}\rightarrow \pi^+\pi^-K^0_SK^0_S$}
&{$55.8\pm5.1\pm8.9$}&{$65\pm6\pm12$}&{-}\\
{$\chi_{c1}\rightarrow \pi^+\pi^-K^0_SK^0_S$}
&{$6.7\pm2.6\pm1.1$}&{$8.0\pm3.1\pm1.5$}&{-}\\
{$\chi_{c2}\rightarrow \pi^+\pi^-K^0_SK^0_S$}
&{$20.7\pm3.9\pm3.3$}&{$32.4\pm6.1\pm6.2$}&{-}\\\hline

{$\chi_{c0}\rightarrow K^+K^-K^0_SK^0_S$}
&{$13.8\pm3.9\pm2.5$}&{$16.0\pm4.6\pm3.2$}&{-}\\
{$\chi_{c1}\rightarrow K^+K^-K^0_SK^0_S$}
&{$2.1\pm1.6\pm0.4$}&{$2.5\pm1.9\pm0.5$}&{-}\\
&$<4.2$ (CL=90\%)&$<5.1$ (CL=90\%)&\\
{$\chi_{c2}\rightarrow K^+K^-K^0_SK^0_S$}
&{$1.6\pm1.6\pm0.3$}&{$2.6\pm2.4\pm0.5$}&{-}\\
&$<3.5$ (CL=90\%)&$<5.5$ (CL=90\%)&\\\hline\hline

{Channel}&{-}&{$BR(\psi(2S)\rightarrow X)$}&{$BR_{PDG}(\psi(2S)\rightarrow X)$\cite{PDG}}\\
{} & {} &{($10^{-4}$)}&{($10^{-4}$)}\\\hline
{$\psi(2S)\rightarrow \pi^+\pi^-K^0_SK^0_S$}
&{-}&{$2.20\pm0.25\pm0.37$}&{-}\\\hline
\end{tabular}
\end{center}
\end{table*}

Decay rates, determined using updated $\chi_{cJ}$ total
widths~\cite{PDG} and branching fractions for $\chi_{cJ}\rightarrow
\pi^0\pi^0,~\pi^+\pi^- ~(J=0,2)$ and $\chi_{cJ}\rightarrow 
p\overline{p}~(J=1,2)$ decays~\cite{PDG}, provide
support for the COM (see Table~\ref{tab:BES data and COM}).
According to isospin symmetry, the $\chi_{cJ} \rightarrow
K^0\overline{K^0}$ and $K^+K^-$ decays should have the same partial
width.  Assuming equal decay widths for $\chi_{cJ} \rightarrow
K_S^0K_S^0$ and $K_L^0K_L^0$, we find that the partial width of the
$\chi_{c0} \rightarrow K^0\overline{K^0}$ decay estimated using the
result obtained in this paper is not consistent (2.7$\sigma$) with the COM
prediction for $\chi_{c0} \rightarrow K^+K^-$, while the agreement between
them for the corresponding $\chi_{c2}$ decay is within 1.1$\sigma$.  
%A similar conclusion can be drawn from 
A comparison for the $\chi_{cJ} \rightarrow K^+K^-~(J=0,2)$ decays
shows that the discrepancy between PDG values
and the COM predictions 
is 2.2$\sigma$ and 1.9$\sigma$ 
for $\chi_{c0}$ and $\chi_{c2}$ decays, respectively.

%This shows that the predictions using the COM for two-body
%final states containing strange mesons may not be correct.

Furthermore, the sum of all known $\chi_{c0}$ two-body branching
fractions is less than $2\%$.  It therefore is important to 
measure more $\chi_{cJ}$ decay modes, including 
two-body modes with intermediate resonance and many-body modes,
%but 
%also to determine the COM contribution for the $\chi_{cJ}$ many-body
%decays 
because of their large contribution to the hadronic decay
width. Theoretical predictions with inclusion of the COM for
$\chi_{cJ}$ decays to many-body final states are required for
comparison with data.

%acknowledgement
\begin{acknowledgments}
The BES collaboration thanks the staff of BEPC for their hard efforts
and the members of IHEP computing center for their helpful assistance,
and also K. T. Chao and J. X. Wang for helpful discussions on the COM.
This work is supported in part by the National Natural Science
Foundation of China under contracts Nos. 19991480,10225524,10225525,
the Chinese Academy of Sciences under contract No. KJ 95T-03, the 100
Talents Program of CAS under Contract Nos. U-11, U-24, U-25, and the
Knowledge Innovation Project of CAS under Contract Nos. U-602,
U-34(IHEP); by the National Natural Science Foundation of China under
Contract No.10175060(USTC),and No.10225522(Tsinghua University); and
by the Department of Energy under Contract No.DE-FG02-04ER41291 (U
Hawaii).\par
\end{acknowledgments}

\begin{table}[t]
\begin{center}
\caption{Comparison of partial widths for $\chi_{cJ}\rightarrow
\pi\pi, K\overline{K}$ and $p\overline{p}$ decays between
PDG~\cite{PDG} and the COM predictions.  Also shown is the result
based on
this analysis.}
\label{tab:BES data and COM}
\begin{tabular}{|c|c|c|}\hline
Decay                               &  $\Gamma_{i}(PDG)$  & $\Gamma_{i}(COM
)$ \\
                                    &   in KeV/c$^2$      & in KeV/c$^2$
   \\\hline
$\chi_{c0}\rightarrow \pi^+\pi^-$   & $49.5\pm6.7$        & 45.4~\cite{JBolz}       \\
$\chi_{c2}\rightarrow \pi^+\pi^-$   & $3.73\pm0.64$       & 3.64~\cite{JBolz}       \\\hline
$\chi_{c0}\rightarrow \pi^0\pi^0$   & $25.3\pm3.3$        & 23.5~\cite{JBolz}       \\
$\chi_{c2}\rightarrow \pi^0\pi^0$   & $2.3\pm1.5$         & 1.93~\cite{JBolz}       \\\hline
$\chi_{c1}\rightarrow p\overline{p}$& $0.066\pm0.015$     & 0.05627~\cite{SMH Wong}\\
$\chi_{c2}\rightarrow p\overline{p}$& $0.143\pm0.018$     & 0.15419~\cite{SMH Wong}\\\hline
$\chi_{c0}\rightarrow K^+K^-$        & $61\pm10$          & 38.6~\cite{JBolz}    \\
$\chi_{c2}\rightarrow K^+K^-$        & $1.98\pm0.47$      & 2.89~\cite{JBolz}    \\\hline
$\chi_{c0}\rightarrow K^0\overline{K^0}$ & $71\pm12$~(this paper)    &       \\
$\chi_{c2}\rightarrow K^0\overline{K^0}$ & $3.76\pm0.80$~(this paper)&       \\\hline
\end{tabular}
\end{center}
\end{table}

\end{document}